\documentclass[aps,twocolumn,superscriptaddress]{revtex4}
\usepackage{amsmath}
\usepackage{color}
\usepackage{bbm}
\usepackage{amssymb}
\usepackage{epsfig}
\usepackage{multirow}
\usepackage{amsbsy}
\usepackage{array}
\usepackage{diagbox}
\usepackage{bm}
\usepackage{tikz}
\usepackage{extarrows}
\usepackage{graphicx}
\usepackage{appendix}
\usepackage{txfonts}
\usepackage[normalem]{ulem}
\usepackage[colorlinks=true,linkcolor=blue,citecolor=blue,urlcolor=blue,bookmarks=false]{hyperref}

\usepackage{soul}

\definecolor{orange}{rgb}{1,0.5,0}

\begin{document}
\title{Dirac and chiral spin liquids on spin-1/2 square-lattice Heisenberg antiferromagnet}

\author{Hui-Ke Jin} 
\email{jinhk@shanghaitech.edu.cn}
\affiliation{School of Physical Science and Technology, ShanghaiTech University, Shanghai 201210, China}

\author{Hong-Hao Tu}
\affiliation{Faculty of Physics and Arnold Sommerfeld Center for Theoretical Physics, Ludwig-Maximilians-Universit\"{a}t M\"unchen, 80333 Munich, Germany}

\author{Ya-Hui Zhang}
\affiliation{William H. Miller III Department of Physics and Astronomy, Johns Hopkins University, Baltimore, Maryland, 21218, USA}

\date{\today}

\begin{abstract}
We revisit the challenging problem of identifying the quantum spin liquid candidate in the spin-1/2 $J_1$-$J_2$ Heisenberg antiferromagnet on the square lattice. By integrating the Gutzwiller-guided density matrix renormalization group method with analytical analyses, we present clear evidence that the ground state is a Z$_2$ Dirac spin liquid. This state can be efficiently described by a Gutzwiller-projected parton theory characterized by its projective symmetry group. To distinguish the difference between the projected Z$_2$ and U(1) parton state, we investigate the chiral spin liquid ground states as topological orders by incorporating a $J_\chi$ term into the $J_1$-$J_2$ model and observe a transition from a Z$_2$ chiral spin liquid to a U(1)$_2$ chiral spin liquid as $J_\chi$ increases.
\end{abstract}

\maketitle

{\em Introduction---}The discovery of high-$T_c$ superconductivity has motivated significant interest in the research on quantum spin liquids (QSL)~\cite{Anderson73,Anderson87,Wen89,Read91,Anderson04,Lee2006}. Over the past few decades, the field of QSL has garnered significant attention, driven by the discovery of emergent long-range entanglement, fractional quasiparticles, and gauge fields~\cite{Lee2008,Balents2010,Savary2016,Zhou17,Knolle2019,Broholm2020}. One of the prototypical and concrete models for QSL is the spin-1/2 $J_1$-$J_2$ antiferromagnetic (AFM) Heisenberg model on the square lattice (hereafter referred to as the $J_1$-$J_2$ model), where $J_1$ and $J_2$ represent the AFM exchanges on the first and second nearest neighbor (NN) bonds, respectively.

It is generally accepted that the $J_1$-$J_2$ model exhibits a paramagnetic phase in the intermediate regime $J_2\approx{}0.5J_1$, where the interplay of quantum fluctuation and geometric frustration destroys the long-range N\'eel order at small $J_2$. However, despite tremendous endeavors by various approaches~\cite{Chandra1988,Dagotto1989,Sachdev1990,Poilblanc1991,Chubukov1991,Schulz1992,Singh1999,Zhitomirsky1996,Capriotti2000,Takano2003,Mambrini2006,Darradi2008,Arlego2008,Yu2012,Doretto2014,Gong2014,Morita2015,Haghshenas2018,Capriotti2001, Wang2013, Hu2013, Poilblanc2017, Wang2018,Liu2018,Ferrari2019,Ferrari2020,Nomura2021,Liu2022,Liu2024,Jiang2012,Qin2024,Isaev2009,Figueirido1990,Ivanov1992,Einarsson1995,Ziman1996,Zhang2003,Sirker2006,Schmalfuss2006,Murg2009,Beach2009,Richter2010,Mezzacapo2012,Qi2014,Chou2014,Merino2022,Richter2015,Wang2016,Poilblanc2019}, the ground state of this paramagnetic phase has been elusive. Possible candidates include (columnar or plaquette) valence-bond solid (VBS)~\cite{Dagotto1989,Sachdev1990,Poilblanc1991,Chubukov1991,Schulz1992,Singh1999,Zhitomirsky1996,Capriotti2000,Takano2003,Mambrini2006,Darradi2008,Arlego2008,Yu2012,Doretto2014,Gong2014,Morita2015,Haghshenas2018}, gapless QSL~\cite{Capriotti2001, Wang2013, Hu2013, Poilblanc2017, Wang2018,Liu2018,Ferrari2019,Ferrari2020,Nomura2021,Liu2022,Liu2024}, and gapped QSL~\cite{Jiang2012}. Besides, a recent study suggests the absence of such a paramagnetic phase~\cite{Qin2024}. Overall, the nature of the intermediate $J_2$ regime remains highly debated.

\begin{figure*}
\includegraphics[width=1\linewidth]{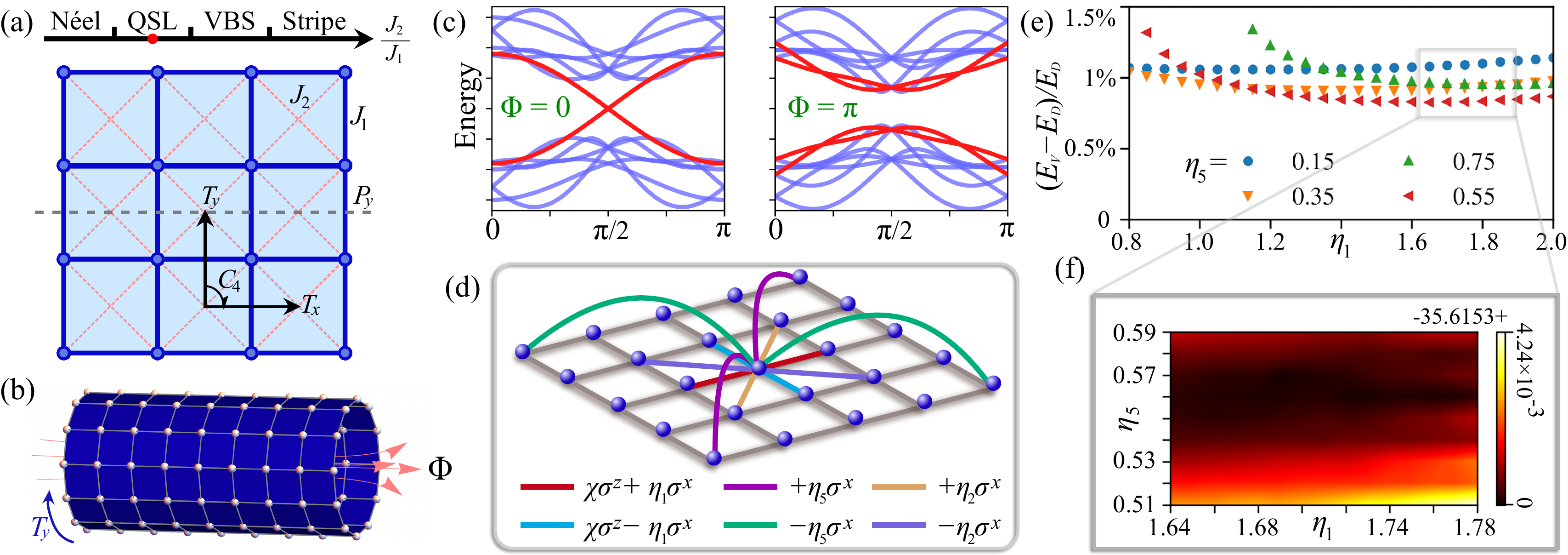}
\caption{The $J_1$-$J_2$ model and $Z_2$ Dirac QSL. 
    (a) Lattice symmetries and ground-state phase diagram of the $J_1$-$J_2$ model on the square lattice. The red dot denotes the point $J_2/J_1=0.5$  being of interest here.
    (b) The square lattice is wrapped onto a cylinder, where the $\hat{y}$-direction is periodic, and the $\hat{x}$-direction remains open. The partons experience an emergent global gauge flux $\Phi$, which effectively introduces a phase factor of $e^{i\Phi}$ for the partons on the boundary.
    (c) For the Z$_2$DSL ansatz on a YC8 cylinder ($\eta_2=0$), the Dirac cones at $\pm(k_y/2, k_y/2)$   can manifest only when $\Phi=0$, while they are gapped a $\Phi=\pi$ flux.
    (d) The projective symmetry group for the $Z_2$ Dirac QSL with the 1$^{\rm st}$,  2$^{\rm nd}$, and 5$^{\rm th}$ NN parton parameters; see Eq.~\eqref{eq:Z2DSL}.  
    (e,f) The variational energy $E_{V}$ for Z$_2$DSL on a YC6 cylinder with length $L_x=12$ with respect to the DMRG-obtained energy $E_{D}\approx{}-35.91$ ($\chi_{D}=8000$). 
    The contour plot (f) provides a zoom-in view of the variational energy landscape. For $\eta_2=0$, the best $E_{V}\approx-35.615$ is obtained at $(\eta_1,\eta_5)=(1.7,0.57)$ with $\chi_{S}=4000$. Further optimizing  $\eta_{2}$ leads to $E_{V}\approx-35.632$.  }\label{fig:fig1}
\end{figure*}

In this work, we revisit this challenging problem and clarify the nature of ground states in the $J_1$-$J_2$ model. We use the density matrix renormalization group (DMRG) method~\cite{white1992,white1993,AnnalPhysSchollwock} and analytical analyses to unveil a  Dirac QSL ground state. 
The key observation is that such a Dirac QSL indeed can be described by a Gutzwiller-projected Z$_2$ parton state within the framework of projective symmetry group (PSG)~\cite{Wen2002}. We emphasize that our numerical results, based on the newly developed method~\cite{Jin2021}, demonstrate an efficient approach to mitigating finite-size effects generally encountered in DMRG simulations. The Z$_2$ QSL also indicates a possible Z$_2$ chiral spin liquid (CSL) phase in the vicinity of the $J_1$-$J_2$ model. The topological order of this Z$_2$ CSL is distinct from the Kalmeyer-Laughlin state~\cite{Kalmeyer87,Kalmeyer89}.
To explore such a potential topological order, we incorporate a spin chirality term $J_\chi$ into the $J_1$-$J_2$ model. As $J_\chi$ increases, we observe a transition from the Z$_2$ CSL phase to a usual U(1) CSL phase.

{\em Model and setup---}
We study the spin-1/2 $J_1$-$J_2$ model on the square lattice, defined as 
\begin{equation}
\mathcal{H}=J_1\sum_{\langle{}ij\rangle_1}\vec{S}_i\cdot\vec{S}_j + J_2\sum_{\langle{}ij\rangle_2}\vec{S}_i\cdot\vec{S}_j, \label{eq:model}
\end{equation}
where $\vec{S}_i$ is the vector of three spin-1/2 operators and $\langle{}ij\rangle_{1(2)}$ runs over all the 1$^{\rm st}$ (2$^{\rm nd}$) NN bonds. The ground-state phase diagram of the $J_1$-$J_2$ model has been extensively explored in previous works~\cite{Jiang2012,Gong2014,Wang2018,Liu2022}, as sketched in Fig.~\ref{fig:fig1}(a). In this study, we focus on the paramagnetic regime supporting QSLs by restricting the model parameters to $J_2=0.5J_1$, corresponding to the maximally frustrated point in the classical limit.

The space group (SG) symmetry of the $J_1$-$J_2$ model is pivotal for the following analysis. As illustrated in Fig.~\ref{fig:fig1}(a), the SG is generated by two translations $T_x$ and $T_y$, a $C_4$ lattice rotation around the out-of-plane $z$-axis, and a parity symmetry $P_y$ (reflection) across the $x$-axis. Besides the SG symmetries, Eq.~\eqref{eq:model} is also invariant under the time-reversal symmetry (TRS) $\mathcal{T}^2=1$ (for an even number of spins). Combining $\mathcal{T}$ with the SG yields the entire symmetry group for the $J_1$-$J_2$ model; see Supplementary Material~\cite{appendix} for further details.

To demonstrate the $Z_2$ Dirac QSL we introduce fermionic parton operators $f^{}_{i,s}$, with $s=1,2$ being the spin up and down index. Then, spin-1/2 operators can be written in a fermion bilinear form $S^a_i=\frac{1}{2}\sum_{s,s'}f^\dagger_{i,s}\sigma^a_{ss'}f^{}_{i,s'}$, where $\sigma^a$ ($a=x,y,z$) are three Pauli matrices. 
This parton representation has an SU(2) gauge redundancy, as straightforwardly evidenced in the framework of Nambu spinor $\psi^{}_i=(f^{}_{i,\uparrow}, f^\dagger_{i,\downarrow})^T$. 
With simple algebra~\cite{appendix}, one can show that the spin operators are invariant under arbitrary SU(2) gauge rotations $\psi^{}_i \rightarrow G^{}_i\psi^{}_i$, where $G_i$ is a {\em local} SU(2) transformation.
This redundancy indicates an enlarged Hilbert space with unphysical double- and non-occupancy states.  The physical Hilbert space is restored by imposing a local single-occupancy constraint, $\sum_{s}f^{\dagger}_{i,s}f^{}_{i,s}=1$. In practice, this procedure, known as Gutzwiller projection, is carried out by exactly projecting out unphysical states, thereby eliminating redundant gauge fluctuations.

This parton representation allows us to systematically construct QSL states~\cite{Wen2002} using the following parton mean-field Hamiltonian:
\begin{equation}
H_{MF}=\sum_{ij}\left(\psi^{\dagger}_i\mu^{}_{ij}\psi^{}_j + \mathrm{h.c.}\right).
\label{eq:HFM}
\end{equation}
Here the mean-field ansatz $\mu_{ij}=\mu_{ji}^\dagger$, which fully characterizes the physical properties of QSL, takes a general form of  $\mu_{ij}=i\chi'_{ij}\sigma^0+\eta_{ij}\sigma^x+\eta^\prime_{ij}\sigma^y + \chi_{ij}\sigma^z$ with $\chi$'s and $\eta$'s being real parameters. Note that, due to the SU(2) gauge structure, the hopping terms of partons ($\sigma^z$ and $\sigma^0$ terms) can be transformed into the singlet pairings of partons ($\sigma^x$ and $\sigma^y$ terms) by a certain gauge transformation, and vice versa. 
The QSL state as a many-body spin wave function is obtained by applying Gutzwiller projection to the parton ground state of Eq.~\eqref{eq:HFM}. 
Because of the SU(2) gauge redundancy in parton representation, the form of mean-field ansatz $\mu_{ij}$ is gauge-dependent and is transformed under gauge rotations as $\mu_{ij}\rightarrow{}G^{}_i\mu_{ij}G^\dagger_j$~\cite{Wen2002}. It indicates that, after exactly implementing the Gutzwiller projection, two seemingly different ansatzs $\mu_{ij}$ and $\mu_{ij}'$ can lead to the same QSL state if there exists a set of $\{G_i\}$ such that $\mu_{ij}=G^{}_i\mu'_{ij}G^\dagger_{j}$. 
Moreover, the form of a symmetric ansatz is allowed to vary under symmetry transformations in SG --- as long as the transformed ansatz is gauge equivalent to its original form~\cite{appendix}.
Thus, when characterizing a QSL, one must also specify its projective symmetries within the framework of PSG~\cite{Wen2002}.

{\em Z$_2$ parton ansatz---}
As a key finding, we report that  the QSL phase in the $J_1$-$J_2$ model is a Z$_2$ Dirac QSL described by the mean-field ansatz as illustrated in Fig.~\ref{fig:fig1}(d)
\begin{equation}
\begin{array}{ll}
    \mu_{i,i+\hat{x}} = \chi{}\sigma^z+\eta_{1}\sigma^x, &
    \mu_{i,i+\hat{y}} = \chi{}\sigma^z-\eta_{1}\sigma^x\\
    \mu_{i,i+2(\hat{x}+\hat{y})} = \eta_{5}\sigma^x,&
    \mu_{i,i+2(\hat{x}-\hat{y})} = -\eta_{5}\sigma^x,\\
      \mu_{i,i+\hat{x}+\hat{y}} = \eta_{2}\sigma^x,&
      \mu_{i,i+\hat{x}-\hat{y}} = -\eta_{2}\sigma^x,
\end{array}~\label{eq:Z2DSL}
\end{equation}
where we fix $\chi_1\equiv\chi=1$ and treat $\eta_{1}$, $\eta_{2}$, and $\eta_{5}$ on the 1$^{\rm st}$, 2$^{\rm nd}$, and 5$^{\rm th}$ NN bonds as variational parameters, respectively.
The mean-field Hamiltonian in Eq.~\eqref{eq:Z2DSL}, in addition to its Z$_2$-type gauge fluctuations, manifests Dirac-type excitations at momenta $\mathbf{k}=(\pm\pi/2,\pm\pi/2)$ when $\eta_2=0$, while a finite value of $\eta_2$ shifts the Dirac cones away form the commensurate point. Thus, we denote this ansatz as Z$_2$DSL.
The PSG solution for the Z$_2$DSL is detailed in the Supplementary Material~\cite{appendix}. 
{In accordance with the PSG, one can find that this Z$_2$DSL turns into a U(1) Dirac QSL for $\eta_{5}=\eta_{2}=0$~\cite{appendix}, wherein Eq.~\eqref{eq:Z2DSL} is equivalent to the staggered-flux state given by~\cite{Lee2006,Affleck1988}, 
\begin{equation*}
    H_{\rm SF} = \sum_{\langle{}ij\rangle_1,s}\exp[i(-1)^{i_x+j_y}\Theta]f^\dagger_{i,s}f^{}_{j,s}+{\rm h.c.},
\end{equation*}
where $\tan\Theta=\eta_1/\chi_1$ and $i_a$ ($a=x, y$) is the lattice index of site $i$ in $a$ direction. Furthermore, the staggered-flux state preserves the SU(2) gauge symmetry at the special point of $\eta_1=\chi_1$. 
It indicates that if either $\eta_2$ or $\eta_5$ is zero, the Z$_2$DSL ansatz with $\eta_1=\chi_1$ actually corresponds to a U(1) Dirac QSL rather than a Z$_2$ one. 
The importance of $\eta_2$ for a Z$_2$ QSL has been investigated previously~\cite{Mudry1994,Shackleton2021}, as the $\eta_2$ term can be naturally anticipated from a straightforward mean-field analysis of the $J_1$-$J_2$ model. Rather than $\eta_2$, later we will see that the $\eta_5$ term, which is unexpected from a mean-field decoupling perspective, plays a more important role in the $J_1$-$J_2$ model.
Furthermore, the PSG solution also implies that other possible terms like $i\sigma^0$ and $\sigma^y$ on the bonds along the $\hat{x}$-, $\hat{y}$-, and $\hat{x}\pm\hat{y}$-directions are forbidden due to the parity symmetries.


The energetics and certain correlation functions of Z$_2$DSL have already been explored using variational Monte Carlo~\cite{Capriotti2001, Hu2013,Ferrari2020}. Here, we shall adopt newly developed methods based on matrix product state (MPS)~\cite{Tu2020,Jin2020,Jin2021,Aghaei2020,Gabriel2021,Jin2022}, which allow us to compute wave function fidelity and entanglement entropy. To perform MPS calculations, we wrap the square lattice onto a finite-circumference cylinder and impose a periodic boundary condition (PBC) on the spin-1/2 variables, as shown in Fig.~\ref{fig:fig1}(b). The cylinder is denoted as YC$L_y$, where $L_y$ represents the circumference, and its length along the open boundary direction is $L_x$. Strictly speaking, this cylindrical geometry preserves the parity symmetry $P_y$ but breaks the $C_4$ symmetry into a twofold $C_2$ rotation symmetry.

Due to the PBC for the spin variables, the fermionic partons placed on cylinders are coupled to an emergent global gauge flux $\Phi$ threading through the cylinder [see Fig.~\ref{fig:fig1}(b)]. This global flux effectively twists the boundary condition for the partons along the $\hat{y}$ direction, i.e., shifting the allowed momenta of partons to $k_y=(2n\pi+\Phi)/L_y$, $n=1,\ldots,L_y$. The product of parity and rotation symmetries $P_yC_2$ requires $\Phi=0$ and $\Phi=\pi$, corresponding to periodic and antiperiodic boundary conditions, respectively. 

In the thermodynamic limit, the effect of $\Phi$ on local observables vanishes as $L_y\rightarrow\infty$.  
However, the global flux $\Phi$ plays an important role in cylindrical systems, as it corresponds to different crystal momenta cutting through the Brillouin zone.  By fixing the gauge for the Z$_2$DSL as in Eq.~\eqref{eq:Z2DSL} (without loss of generality we make $\eta_2=0$), the allowed momenta exactly cut the Dirac cones of Z$_2$DSL at $\pm(\pi/2, \pi/2)$ only when $\Phi=0$ ($\Phi = \pi$) for $L_y/2$ being even (odd). For instance, the one-dimensional band structure for Z$_2$DSL on a YC8 cylinder is illustrated in Fig.~\ref{fig:fig1}(c), where it exhibits a (finite-size) band gap for $\Phi=\pi$, in contrast to the appearance of Dirac cones for $\Phi=0$.
It indicates that even if the $J_1$-$J_2$ model is described by a gapless Z$_2$DSL ansatz, it may manifest spurious characteristics of gapped states when placed on certain types of finite cylinders. Consequently, one may misidentify the $J_1$-$J_2$ model as a gapped QSL or other states when focusing only on local quantities like energetics and (short-distance) correlation functions.

{\em Numerical results---}
Apart from the parton ansatz, we also employ the DMRG method to simulate the $J_1$-$J_2$ model directly. The U(1) quantum number corresponding to $S^z$ conservation is explicitly used, with bond dimensions for preparing MPS and performing DMRG calculations denoted by $\chi_{S}$ and $\chi_{D}$, respectively. The DMRG calculations are initialized with randomly generated MPSs or Gutzwiller projected parton states~\cite{Jin2021}.

We treat the global flux $\Phi=0$ and $\pi$ as a discrete parameter and compare the variational energy $E_{V}$ of Z$_2$DSL in both flux sectors. For the same variational parameters, we find that the lower-energy sector is always the one wherein the Dirac cones are avoided. 
By filling the single-particle energy levels below the Dirac cones, the ``gapped'' sector gains energy of $\delta E_V\sim{}v_F/L_y$ by avoiding cutting the Dirac cones, where $v_F$ is the spinon Fermi velocity.  
Note that this scenario also explains why the singlet energy gap in the $J_1$-$J_2$ model scales as $L_y^{-1}$~\cite{Wang2018,Qin2024}, since the parton band gap introduced by $\Phi$ closes proportionally as $v_F\delta{}k_y{}\propto{}v_FL_y^{-1}$. 

In the following, we concentrate on the ``gapped'' sector and optimize the parameters $\eta_{1}$ and $\eta_{5}$ of Z$_2$DSL by minimizing the variational energy $E_V$. After a careful search in the parameter space, as demonstrated in Fig.~\ref{fig:fig1}(e-f), we find that the lowest energy $E_V$ is always achieved with a finite $\eta_{5}$.  For instance, the optimal values of $\eta_{5}$ are 0.42, 0.57, and 0.58 for YC4, YC6, and YC8 cylinders, respectively. These values are consistent with variational Monte Carlo results~\cite{Hu2013,Ferrari2020}. 
We find that the quality of the Z$_2$DSL ansatz is not sensitive to $\pm{}\eta_{2}\sigma^x$ terms on the 2$^{\rm nd}$ NN bonds [see Fig.~\ref{fig:fig1}(d)]. This term preserves the PSG of Z$_2$DSL and shifts the Dirac cone away from $(\pm\pi/2,\pm\pi/2)$ (with our gauge choice). Although the energy can be slightly improved with a finite $\eta_{2}\approx0.04$, we set $\eta_{2}=0$ in discussions below for simplicity. 

We emphasize that the presence of finite $\eta_{5}$,  besides contributing to additional energy gain, is also essential for breaking the SU(2) gauge structure down to a Z$_2$ one. As mentioned above, for $\eta_5=\eta_2=0$, Z$_2$DSL is equivalent to the U(1) staggered-flux state.
Nevertheless, our numerics clearly demonstrate that for the $J_1$-$J_2$ model (i) the optimal value of $\eta_5$ ($\eta_2$) is finite (close to zero) and (ii) the optimal value of $\eta_1$ is far away from the special SU(2) point of $\eta_1=\chi_1$, as shown in Fig.~\ref{fig:fig1}(e). Overall, all these results suggest that the $J_1$-$J_2$ model is more likely to support a gapless Z$_2$ QSL rather than a U(1) QSL in the 2D limit.  

We also perform DMRG calculations initialized with randomly generated MPS and Z$_2$DSL ansatz, both of which converge to the same unbiased ground-state energy $E_{D}$. In contrast, utilizing a Z$_2$DSL initial state significantly reduces the wall time required to achieve converged energy by half, in comparison to starting with a random MPS~\cite{appendix}. Moreover, the relative difference between $E_D$ and variational energy $E_V$ is as small as $0.8\%$, as shown in Fig.~\ref{fig:fig1}(e).
The validity of the Z$_2$DSL ansatz can be further confirmed by calculating the wave function fidelity $F = |\langle \Psi_{\text{DMRG}} | \Psi_{Z_2\text{DSL}} \rangle |$. Notably, the per-site fidelity achieves a high value of approximately $$\tilde{f}\equiv{}F^{1/L_xL_y}\approx99.86\%,$$ which turns out to be unchanged with $L_x$ and even $L_y$~\cite{appendix}.
For instance, the fidelity on YC6 and YC8 cylinders with $L_x=12$ read $F\approx0.91$ and $F\approx0.87$, respectively. {\color{black} Remarkably, the per-site $\tilde{f}$ persists at a large value $>99.85\%$ for the entire QSL regime of $0.45\leq{}J_2\leq{}0.5$, reaching its maximal at $J_2\approx0.47$~\cite{appendix}. 
We stress that this Z$_2$DSL is robust under perturbations such as a VBS ordering and N\'eel ordering~\cite{appendix}.}
We further examine the SU(2) $\pi$-flux ansatz~\cite{Wen2002} which also exhibits Dirac spinon excitations. However, our results are not in favor of such an SU(2) $\pi$-flux state as the candidate ground state of the $J_1$-$J_2$ model~\cite{appendix}. 
Overall, we provide clear evidence that the ground state of the \( J_1\text{-}J_2 \) model is indeed a Z$_2$ Dirac QSL described by our parton ansatz.

\begin{figure}
\includegraphics[width=\linewidth]{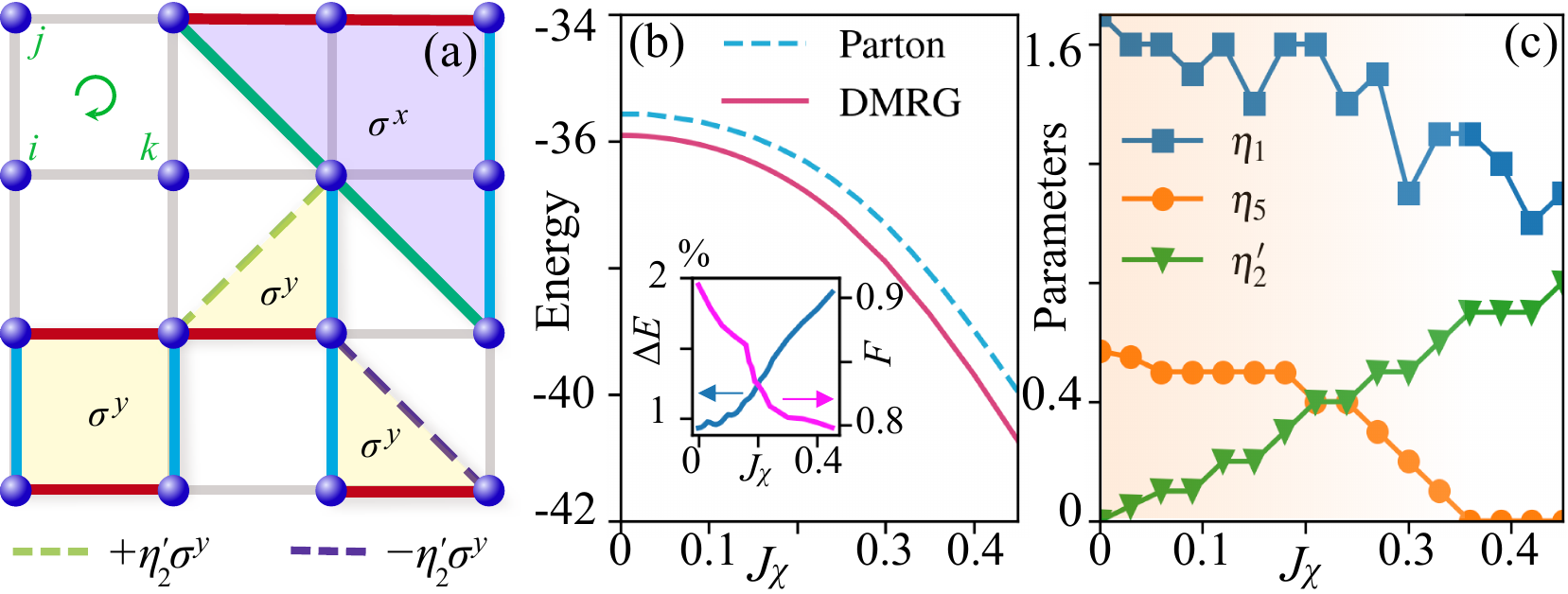}
\caption{The Z$_2$ and U(1) CSLs in the $J_1$-$J_2$-$J_\chi$ model. (a) The sites $i$, $j$, and $k$ indicate the chiral terms discussed in Eq.~\eqref{eq:HJx}. The dashed off-diagonal lines denote $\eta_{2}'$ terms in Eq.~\eqref{eq:eta2p}. The enclosed flux within the elementary square, elementary triangular, and big triangular plaquettes sustains U(1), U(1), and Z$_2$ gauge fluctuations, respectively.  (b)  The best variational parton energy ($\chi_{S}$=2000) and DMRG energy ($\chi_{D}$=4000) as functions of $J_\chi$ on a YC6 cylinder with $L_x=12$. Inset: The relative energy difference (left axis), $\Delta{}E=(E_D-E_V)/E_D$ slowly increases with $J_\chi$. The wave function fidelity (right axis), $F=|\langle \Psi_{\text{DMRG}} | \Psi_{\text{CSL}} \rangle |$, as a function of $J_{\chi}$. (b) The optimized parameters as functions of $J_\chi$. For $J_\chi\gtrsim0.35$, $\eta_{5}$ vanishes, and the system undergoes a transition from a Z$_2$ CSL to a U(1) CSL. }\label{fig:fig2}
\end{figure}

{\em CSLs by spin chirality---}
The presence of Z$_2$DSL suggests that incorporating a chiral term into the $J_1$-$J_2$ model can break time-reversal symmetry and lead to an unusual Z$_2$ CSL ground state. To this end, we consider the $J_1$-$J_2$-$J_\chi$ model
\begin{equation}
\mathcal{H}_\chi=\mathcal{H}+J_{\chi}\sum_{ijl\in{}\triangle}\vec{S}_i\cdot(\vec{S}_j\times{}\vec{S}_{k}),\label{eq:HJx}
\end{equation}
where the sites $i$, $j$ and $k$ in each triangle $ijl\in{}\triangle$ follow the same clockwise direction [see Fig.~\ref{fig:fig2}(a)]. Besides the $\mathcal{T}$ symmetry, the lattice parity symmetry $P_y$ is also broken explicitly by the chiral $J_{\chi}$ term. Nevertheless, the rotation symmetry $C_4$ is preserved in the presence of $J_{\chi}$. This chiral term can gap out the Dirac cones in Z$_2$DSL~\cite{Nielsen2013,Poilblanc2017Chiral,Hasik2022,Zhang2024,Yang2024}, leading to gapped CSL ground states characterized by topological order~\cite{Wenbook}. 

When $J_{\chi}$ is relatively small, one can intuitively expect that the Z$_2$DSL ansatz incorporating TRSB perturbations could be a promising candidate for the ground state of Eq.~\eqref{eq:HJx}.  
Thus, several additional terms, which were ruled out by the PSG of Z$_2$DSL~\cite{appendix}, should be taken back as the perturbations for the Z$_2$DSL. Thus, apart from the terms in Eq.~\eqref{eq:Z2DSL}, we introduce an additional set of parameters for the CSL regime:
\begin{subequations}
\begin{eqnarray}
&&\mu'_{i,i+\hat{x}}=i\chi^\prime_1\sigma^0+\eta^\prime_{1}\sigma^y,\quad
    \mu'_{i,i+\hat{y}}=-i\chi^\prime_1\sigma^0-\eta^\prime_{1}\sigma^y,\\
&&\mu'_{i,i+\hat{x}\pm\hat{y}}=i\chi^\prime_2\sigma^0\pm\eta^\prime_{2}\sigma^y.\label{eq:eta2p}
\end{eqnarray}
\end{subequations} 
In the gauge we have chosen, $\eta'_1$ and $\eta'_2$ are the $id_{x^2-y^2}$ and $id_{xy}$ pairings on the 1$^{\mathrm{st}}$ and 2$^{\mathrm{nd}}$ NN bonds, respectively.  
After a careful investigation, we find that only the $\eta_2'$ term significantly improves the variational energy with varying $J_\chi$~\cite{appendix}. Therefore, only $\eta_2'$ among the aforementioned parameters is pivotal in mimicking the low-energy physics of the $J_1$-$J_2$-$J_\chi$ model. Indeed, $\eta_2'$ is an effective mass term for the Dirac fermions in the Z$_2$ Dirac QSL. Consequently, we proceed to investigate potential CSL phases by focusing on the CSL ansatz, which is a Gutzwiller projected $d+id$ state with variational parameters $(\eta_1, \eta_2', \eta_5)$.  

For a given value of $J_\chi$, we calculate the ground-state energy of the $J_1$-$J_2$-$J_\chi$ model using DMRG as well as the variational energy of the Gutzwiller projected parton states parametrized by three parameters $(\eta_1, \eta_2', \eta_5)$. We observed that DMRG calculations initialized with parton states and random MPSs give rise to the same converged ground states. As shown in Fig.~\ref{fig:fig2}(b), despite a gradual and slow increase with $J_\chi$,  the relative energy difference always remains below $2\%$ for $J_\chi<0.4$.  We also computed the wave function fidelity $F$ between the state obtained by DMRG and the Gutzwiller projected parton state on a YC6 cylinder with $L_x=12$. 
As $J_\chi$ increases, $F$ undergoes an almost linear decrease from $F\approx0.91$ (for $J_\chi = 0$) to $F\approx0.81$ ($J_\chi = 0.3$). Subsequently, its value slowly decreases or fluctuates around $0.8$ for $J_\chi<0.45$, as shown in the inset of Fig.~\ref{fig:fig2}(b). Overall, In the regime of our interest, the per-site fidelity is always larger than $99.68\%$, indicating that the whole CSL phase can be well described by our CSL parton ansatz, in particular for $J_\chi<0.2$.

We uncover a possible phase transition within the CSL regime, as revealed by the evolution of the optimal parton parameters with  $J_\chi$. As shown in Fig.~\ref{fig:fig2}(c), $\eta_5$ (the $d_{xy}$ pairing on the 5$^{\rm th}$ NN bonds) stays at a plateau value around 0.5 when $J_{\chi}\lesssim{}0.2$,  and subsequently declines to zero as $J_{\chi}\approx0.35$. Meanwhile, $\eta_2'$, representing the $id_{xy}$ pairing on the 2$^{\rm nd}$ NN bonds, exhibits an approximately linear increase with the rise in $J_\chi$.  Finally, $\eta_{1}$ always remains finite on the 1$^{\rm st}$ bond. The vanish of $\eta_{5}$ indicates a phase transition from a Z$_2$ CSL to a U(1) CSL. To see this, we carefully examine the nonzero SU(2) gauge fluxes~\cite{Wen2002, appendix} in our CSL ansatz.  As shown in Fig.~\ref{fig:fig2}(a), the SU(2) gauge fluxes around an elementary square and/or an elementary triangle manifest as $a\sigma^0+b\sigma^y$, where $a$ and $b$ are complex numbers. Since they are invariant under arbitrary U(1) gauge transformations generated by $\sigma^y$, the set of parameters ($\eta_1$, $\eta_2'$) indeed corresponds to a U(1) CSL. However, the gauge flux around the large triangle [see Fig.~\ref{fig:fig2}(a)] has a form of $\sim{}\eta_5\sigma^x$ and thereby only the Z$_2$ number $\pm{}\sigma^0$ can simultaneously commute the two fluxes proportional to $\sigma^y$ and $\sigma^x$, respectively. Consequently, a finite value of $\eta_{5}$ changes such a U(1) CSL into a Z$_2$ CSL. When the effect of $J_\chi$ is small, the system is in the vicinity of the Z$_2$DSL state, and naturally, it supports a Z$_2$ CSL wherein $\mathcal{T}$ is explicitly broken by the $J_{\chi}$ term. 

We argue that the Z$_2$CSL state is characterized by the $\nu=4$ state in Kitaev's sixteenfold way of anyon theories~\cite{Kitaev06,Song2021,Chulliparambil2020}, with a chiral central charge $c=2$~\cite{appendix}. Nevertheless, the U(1) CSL, which has the same topological order as the Kalmeyer-Laughlin state~\cite{Kalmeyer87,Kalmeyer89}, is not part of the sixteenfold way theories and has a chiral central charge $c=1$. 
This distinction can be directly revealed by the topological ground-state degeneracy on a torus~\cite{Zhang2011,Zhang2012,Zhang2013}. When placing the Gutzwiller projected state on the torus, four states can be constructed by changing the boundary conditions of parton mean-field Hamiltonian in both $\hat{x}$ and $\hat{y}$ directions. In the case of the Z$_2$ CSL, we have found that four states are linearly independent, forming the topologically degenerate ground-state manifold. In contrast, for the U(1) CSL, we confirmed that the ground-state degeneracy is reduced to two, in agreement with the U(1)$_2$ topological order. {\color{black} The details for the possible criticality behavior between the U(1) and Z$_2$ CSLs can be found in the Supplementary Material~\cite{appendix}. }

{\em Summary---}
In conclusion, we revisit the $J_1$-$J_2$ model in the paramagnetic regime by combining the Gutzwiller-guided DMRG method and analytical analyses. By examining the ground-state energies and, particularly, the wave function fidelities, we provide direct and strong evidence that the QSL phase in the  $J_1$-$J_2$ model is a Z$_2$ Dirac QSL. To explore the potential topologically ordered phases in the vicinity of the Z$_2$ Dirac QSL, we add a spin chirality $J_\chi$ term in the $J_1$-$J_2$ model and map out its phase diagram, where we find two distinct CSLs, i.e., Z$_2$ and U(1) CSLs, separated by a phase transition. With the help of PSG analysis, we find that all of the Dirac QSL and two CSLs can be efficiently described by Gutzwiller projected parton states. 
Apart from certain exactly solvable models, our results strongly clarify the validity of the parton construction method in a crucial and concrete model.

In addition to enhancing computational efficiency, the Gutzwiller-guided DMRG method serves as a robust and potent platform for directly evaluating various key properties, e.g., wave function fidelities. 
Gutzwiller projected wave function is a versatile tool for {\color{black} detecting effective parton ansatz~\cite{appendix}, by which a field theory of criticality between different phases in the $J_1$-$J_2$ model can be possibly constructed.}
Meanwhile, this method allows us to 
systematically studying topological order~\cite{Jin2021,Sun2024}. 
In the future, it will be interesting to prepare the parton states in the semion sectors of Z$_2$ CSL by using this method, and then adiabatically evolve it by DMRG~\cite{Sun2024} to capture two different semion sectors in the $J_1$-$J_2$-$J_\chi$ model.

Our research could spark renewed interest in this pendent issue and related problems.
(i) Doping this Z$_2$ Dirac QSL or CSLs to explore possible superconducting states.
(ii) It would be desirable to further explore topological properties such as entanglement spectra for the Z$_2$ CSL phase by DMRG calculations on wider cylinders, variational Monte Carlo on larger tori, and tensor network calculations in the thermodynamic limit.
(iii) Our PSG analysis points out that the parameters on the 4$^{\rm th}$ bonds of the parton Hamiltonian support a CSL ansatz preserving all the lattice symmetry. Whether such a CSL can emerge in the $J_1$-$J_2$-$J_4$ model through spontaneously breaking TRS is worth exploring.

{\em Acknowledgements---}
We thank Francesco Ferrari for the communications on variational Monte Carlo data.
H.-K. Jin acknowledges the support from the start-up funding from ShanghaiTech University. YHZ was supported by a startup fund from Johns Hopkins University and the Alfred P. Sloan Foundation through a Sloan Research Fellowship (YHZ).

\bibliography{QSLJ1J2}

\clearpage
\newpage

\begin{widetext}
\begin{center}
{\bf Supplementary information for \\``Dirac and chiral spin liquids on spin-1/2 square-lattice Heisenberg antiferromagnet''}\\
\vspace{10px}

{Hui-Ke Jin,$^{1}$~~Hong-Hao Tu,$^{2}$,~~Ya-Hui Zhang$^{3}$}\\
\vspace{10px}

{\em 
{$^1$School of Physical Science and Technology, ShanghaiTech University, Shanghai 201210, China}\\
\vspace{1px}
{$^2$Faculty of Physics and Arnold Sommerfeld Center for Theoretical Physics, \\ 
Ludwig-Maximilians-Universit\"{a}t M\"unchen, 80333 Munich, Germany}\\
\vspace{1px}
{$^3$William H. Miller III Department of Physics and Astronomy,\\ Johns Hopkins University, Baltimore, Maryland, 21218, USA}\\
\vspace{10px}
}
\end{center}    

This Supplemental Material includes more details on (i) parton construction and SU(2) flux, (ii) brief projective symmetry group on the square lattice, (iii) quantum spin liquid ansatz for the $J_1$-$J_2$ model, (iv) details on the  $J_1$-$J_2$-$J_\chi$ model, (v) more details on density matrix renormalization group data, {\color{black}(vi) instability against VBS state and N\'eel order, (vii) topological ground-state degeneracy, and (viii) field theory of criticality between the U(1) and Z$_2$ CSLs.}

\vspace{5px}

\end{widetext}

\section{Parton construction and SU(2) flux}

To demonstrate the parton construction clearly, one can use a fermionic parton field
\begin{eqnarray}
\hat{F}_i\equiv\left(\begin{array}{cc}
f^{}_{i\uparrow}  &  -f^\dagger_{i\downarrow}\\
f^{}_{i\downarrow}  &  f^\dagger_{i\uparrow}\\	
\end{array}\right),
\end{eqnarray} 
and fractionalize the spin operators in the representation such that 
\begin{equation}
    S^a_i = \frac{1}{4}\text{Tr}(\hat{F}^\dagger_{i}\sigma^a{}\hat{F}^{}_{i}).
\end{equation} 
It is easy to verify that the above representation can recover the standard Abrikosov fermion representation of spins. 
Any right SU(2) rotation $\hat{F}_i \rightarrow \hat{F}_iG_i$, $G_i \in $ SU(2) leaves the physical spin $S^a_i$ (and hence the spin-spin interactions) unchanged. Therefore the right rotation $G_i$ corresponds to a gauge SU(2) rotation, which generates the SU(2) gauge structure in the parton construction method.

The parton mean-field Hamiltonian in terms of ansatz $u_{ij}$, which has been defined in Eq.~(2) in the main text, can be explicitly written as
\begin{equation}
\begin{split}
H_{MF}=&\sum_{ij,s=\uparrow,\downarrow}\left[(i\chi'_{ij}+\chi_{ij})f^\dagger_{is}f^{}_{js}+\mathrm{h.c.}\right]\\&+\sum_{ij}\left[(\eta_{ij}+i\eta'_{ij})(f^\dagger_{i\uparrow}f^\dagger_{j\downarrow}-f^\dagger_{i\downarrow}f^\dagger_{j\uparrow})+\mathrm{h.c.}\right].
\end{split}
\end{equation}
Two different mean-field Hamiltonians can lead to the same QSL state. We use the parton ansatz defined in Eq.~(3) in the main text as an example. Restricting ourselves to the 1st NN bonds, the parton ansatz (we make $\chi_1=\eta_1=1$) is 
\begin{equation}
\mu_{i,i+\hat{x}} = \sigma^z+\sigma^x, ~
\mu_{i,i+\hat{y}} = \sigma^z-\sigma^x.\label{eq:Gauge1}
\end{equation}
We can choose another gauge convention using the following gauge transformation 
\begin{equation}
G_{i} = \sigma^z. \label{eq:GT1}
\end{equation}
Then, Eq.~\eqref{eq:Gauge1} becomes 
\begin{equation}
\mu_{i,i+\hat{x}} = \sigma^z-\sigma^x, ~
\mu_{i,i+\hat{y}} = \sigma^z+\sigma^x.\label{eq:Gauge2}
\end{equation}
It is not difficult to verify that applying Eq.~\eqref{eq:GT1} to the ansatz in Eq~\eqref{eq:Gauge1} is equivalent to a $C_4$ lattice rotation. It indicates that Eq~\eqref{eq:Gauge1} is $C_4$ rotation invariant.

The type of $H_{MF}$  (such as SU(2), U(1), and Z$_2$) can be characterized by the SU(2) flux defined on a closed loop with a base site~\cite{Wen2002}. 
For instance, denoting $r_1,r_2,r_3,r_4$ as for lattice sites forming a square in clockwise order, we can define a SU(2) flux on the site $r_1$ as 
\begin{equation}
F_{\square}(r_1) = \mu_{r_1,r_2}\mu_{r_2,r_3}\mu_{r_3,r_4}\mu_{r_4,r_1}.
\end{equation}
It is transformed under a SU(2) gauge transformation on $r_1$ as 
\begin{equation}
F_{\square}(r_1) \rightarrow{}G^{}_{r_1}F_{\square}(r_1) G^{\dagger}_{r_1}.
\end{equation}
Therefore, the SU(2) flux does not necessarily commutate with the local gauge transformation. If all nonzero SU(2) fluxes manifest as $F_{\square}\sim{}\sigma^0$, then the SU(2) gauge structure is not broken and the corresponding $H_{MF}$ leads to a SU(2) QSL. If $F_{\square}$ consists of at least one (two) of three Pauli matrices, $H_{MF}$ describes a U(1) (Z$_2$) QSL.

For the Z$_2$DSL, the SU(2) flux enclosed within the elementary square plaquette reads
\begin{equation}
\begin{split}
    F_{\square}(i) &= \mu_{i,i+\hat{x}}\mu_{i+\hat{x},i+\hat{x}+\hat{y}}\mu_{i+\hat{x}+\hat{y},i+\hat{y}}\mu_{i+\hat{y},i}    \\
    &=\left[(\chi_1^2-\eta_1^2)\sigma^0 - 2i\chi\eta\sigma^y\right]^2\sim{}a\sigma^0+b\sigma^y.~\label{eqS:ES}
\end{split}
\end{equation}
Therefore, this flux breaks the SU(2) gauge symmetry into a U(1) one. Note that the above flux $F_{\square}$ is only proportional to $\sigma^0$ when $\chi_1=\eta_1$, namely, at this special point it supports an SU(2) rather than U(1) gauge symmetry. If we consider the bigger triangular plaquette, the corresponding gauge flux reads
\begin{equation}
\begin{split}
    F_{\bigtriangleup}(i) &= \mu_{i,i+\hat{x}}\mu_{i+\hat{x},i+2\hat{x}}\mu_{i+2\hat{x},i+2\hat{x}+\hat{y}}\mu_{i+2\hat{x}+\hat{y},i+2\hat{x}+2\hat{y}}\mu_{i+2\hat{x}+2\hat{y},i}    \\
    &=(\chi_1^2+\eta_1^2)^2\eta_5\sigma^x.
\end{split}~\label{eqS:BT}
\end{equation}
Since now only $\pm{}\sigma^0$ can commutate with Eqs.~\eqref{eqS:ES} and~\eqref{eqS:BT} simultaneously, the presence of Eq.~\eqref{eqS:BT} further reduces the U(1) gauge symmetry into a Z$_2$ one. Importantly, the Z$_2$DSL ansatz with a finite $\eta_{5}$ describes a Z$_2$ QSL.
For the CSL ansatz, the gauge flux enclosed within the elementary triangular plaquette is 
\begin{equation}
\begin{split}
    F_{\triangleleft}(i) &= \mu_{i,i+\hat{x}}\mu_{i+\hat{x},i+\hat{x}+\hat{y}}\mu_{i+\hat{x}+\hat{y},i}    \\    
    &=\left[(\chi_1^2-\eta_1^2)\sigma^0 - 2i\chi_1\eta_1\sigma^y\right]\eta_2^\prime\sigma^y.
\end{split}~\label{eqS:ET}
\end{equation}
Thus Eqs.~\eqref{eqS:ET} and~\eqref{eqS:ES} support the same kind of gauge fluctuations.

\section{PSG on the square lattice}

Before discussing the projective symmetry group (PSG), we shall first clarify the symmetry group of the square lattice.  We use $(x,y)$ to denote the site of the square lattice. The translations $T_x$ and $T_y$ are applied as 
\begin{equation}
({x,  y}) \xrightarrow{T_x} ({ x+1,  y}), \quad
({x, y}) \xrightarrow{T_y} ({ x,  y+1}).
\end{equation}
The two parity symmetries, $P_{y}$ and $P_{xy}$,  indicate that 
\begin{equation}
({x,  y}) \xrightarrow{{P_y}} ({ x,  -y}),\quad
({x,  y}) \xrightarrow{{P_{xy}}} ({ y,  x}).
\end{equation}
For simplicity of notation, we introduce a  parity symmetry dubbed $P_x=P_{xy}P_yP_{xy}$ which leads to 
\begin{equation}
({x,  y}) \xrightarrow{{P_x}} ({ -x,  y}).
\end{equation}
The 90$^\circ$ rotation symmetry is given as $C_4=P_yP_{xy}$ as 
\begin{equation}
({x,  y}) \xrightarrow{{P_{xy}}} ({ y,  x})\xrightarrow{{P_y}}({y,-x}). 
\end{equation}
The above symmetric transformations lead to the algebraic relations
\begin{subequations}
\begin{eqnarray}
    T_{x}T_{y}T_{x}^{-1}T_y^{-1}&=&1,\\
    (P_y)^2=(P_{xy})^2&=&1,\\
    T_xP_{xy}P_{y}P_{xy}T_xP_{xy}P_{y}P_{xy}&=&1,\\
    T_yP_yT_yP_y&=&1,\\
    T_x^{-1}P_{xy}T_yP_{xy}&=&1,\\
    T_y^{-1}P_{xy}T_xP_{xy}&=&1,\\
    (P_yP_{xy})^4&=&1.
\end{eqnarray}
\end{subequations}
The above equations can give rise to corresponding algebraic PSG constraints~\cite{Wen2002}, namely, 
\begin{equation*}
g_{m},...,g_2g_{1}=1 \Longrightarrow{} G_{g_{m}}[g_{m}^{-1}(i)]...G_{g_2}[g_1(i)]G_{g_1}(i)= G_{\rm IGG},
\end{equation*}
with $G_{g}(i)$ being the local gauge transformation for symmetry transformation $g$ at site $i$, and $G_{\rm IGG}$ is the elements belonging to the invariance gauge group of PSG. For a U(1) QSL ansatz, the $G_{\rm IGG}$ is equivalent to a U(1) group, while for a Z$_2$ QSL ansatz, it is a Z$_2$ group. For a given $G_{\rm IGG}$, one can obtain the algebraic PSG by solving the above algebraic PSG constraints. 
The  parton mean-field parameters are transformed by a symmetric operation $g$ as
\begin{equation}
\mu_{i,j}\xrightarrow{ g } \mathcal{G}[g,\mu_{i,j}]\equiv{}G^{g}(i) \mu_{g(i),g(j)} [G^{g}(j)]^{-1}. \label{eq:GTuij}
\end{equation}
If such a parton ansatz describes a symmetric QSL, then it requires $ \mu_{i,j}\overset{!}{=}\mathcal{G}[g,\mu_{i,j}].$

\begin{table}
\caption{Lattice and projective symmetries for Z$_2$DSL on the square lattice }\label{tab:Z2DSL}
\begin{tabular}{ p{1.2cm} p{3.5cm} p{2.cm} }
    \hline
    \hline
    \rule{0pt}{2.5ex}    
    & symmetry & PSG operation \\
    \hline
    \rule{0pt}{3ex}    
    $T_x$ & $({x,  y}) \rightarrow ({ x+1,  y})$ & $\sigma^0$\\
    $T_y$ & $({x,  y}) \rightarrow ({ x,  y+1})$ & $\sigma^0$\\
    $P_y$ & $({x, y}) \rightarrow ({ x,  -y})$ & $(-1)^{x+y}\sigma^y$\\
    $P_{x}$ & $({x, y}) \rightarrow ({-x,  y})$ & $(-1)^{x+y+1}\sigma^y$\\
    $P_{xy}$ & $({x, y}) \rightarrow ({ y,  x})$ & $(-1)^{x+y}\sigma^x$\\
    $C_4$ & $({x, y}) \rightarrow ({ y,  -x})$ & $-i\sigma^z$\\
    \hline
    \hline
\end{tabular}
\end{table}

By fixing a specific gauge of $G_{T_1}(i)=G_{T_2}(i)=\sigma^0$,  the algebraic PSG supporting a Z$_2$DSL ansatz is summarized in Table.~\ref{tab:Z2DSL}.
Note that we have fixed a SU(2) gauge to make  Eq.~(3) in the main text be translational invariant. 

The PSG of Z$_2$DSL also indicates that the time-reversal-symmetry must be preserved if we only include parton mean-field parameters on the bonds along the $\hat{x}$, $\hat{y}$, and $\hat{x}+\hat{y}$ directions. 
The $P_y$ symmetry requires that 
\begin{equation}
\begin{split}
&\mu_{i,i+\hat{x}}=\mathcal{G}[P_y, \mu_{i,i+\hat{x}}]\overset{!}{=}-\sigma^y\mu_{i,i+\hat{x}}\sigma^y,\\
&\mu_{i,i+2\hat{x}+2\hat{y}}=\mathcal{G}[P_y, i+2\hat{x}+2\hat{y}]\overset{!}{=}\sigma^y\mu_{i,i-2\hat{x}+2\hat{y}}\sigma^y.~\label{eqS:PSGPyZ2DSL}
\end{split}
\end{equation}
It is easy to verify that Eq.~(3) in the main text satisfies Eq.~\eqref{eqS:PSGPyZ2DSL}. Note that the above constraint equations also indicate that the terms of $i\chi_{1}^\prime\sigma^0+\eta_1^\prime\sigma^y$ should vanish in Z$_2$DSL ansatz because of the $P_y$ symmetry. 
Analogous constraint equations to Eq.~\eqref{eqS:PSGPyZ2DSL} can be derived for the $P_x$ and $P_{xy}$ symmetries. For instance, the PSG of Z$_2$DSL requires that 
\begin{equation}
\begin{split}
&\mu_{i,i+2\hat{x}+2\hat{y}}=\mathcal{G}[P_{xy},\mu_{i,i+2\hat{x}+2\hat{y}}]=\sigma^x\mu_{i,i+2\hat{x}+2\hat{y}}\sigma^x,\\
&\mu_{i,i+2\hat{x}+2\hat{y}}=\mathcal{G}[P_{y},\mu_{i,i+2\hat{x}+2\hat{y}}]=\sigma^y(\mu_{i,i-2\hat{x}+2\hat{y}})^\dagger\sigma^y,\\
&\mu_{i,i+2\hat{x}+2\hat{y}}=\mathcal{G}[P_{x},\mu_{i,i+2\hat{x}+2\hat{y}}]=\sigma^y\mu_{i,i-2\hat{x}+2\hat{y}}\sigma^y.	\end{split}~\label{eqS:PSGPxyZ2DSL}
\end{equation}
Then a straightforward calculation reveals that the time-reversal-symmetry breaking terms such as $i\chi_{5}^\prime\sigma^0+\eta_5^\prime\sigma^y$ always vanishes on the bonds along the $\hat{x}\pm{}\hat{y}$ directions due to the parity symmetries. 

Note that the time-reversal-symmetry-breaking terms on bonds along the other directions are in principle allowed by parity symmetries.  For instance,  we consider the terms on the 4th NN bonds with a general form of $\mu_{i,i+2\hat{x}+\hat{y}}=i\chi_4^\prime\sigma^0+\eta^\prime_4\sigma^y.$ Then the parity symmetries require 
\begin{equation}
\begin{split}
&\mu_{i,i+2\hat{x}+\hat{y}} = \mathcal{G}[P_y, \mu_{i,i+2\hat{x}+\hat{y}}]\overset{!}{=}-\sigma^y(\mu'_{i,i-2\hat{x}+\hat{y}})^\dagger\sigma^y,	\\
&\mu_{i,i+2\hat{x}+\hat{y}} = \mathcal{G}[P_x, \mu_{i,i+2\hat{x}+\hat{y}}]\overset{!}{=}-\sigma^y\mu'_{i,i-2\hat{x}+\hat{y}}\sigma^y.	\\
\end{split}
\end{equation}
Though the imaginary hopping $\chi_4^\prime$ should vanish due to the symmetry, the pairing term $\eta_4^\prime$ can be a finite value.

\section{QSL ansatz for $J_1$-$J_2$ model}

The explicit form of the Z$_2$DSL ansatz is 
\begin{equation}
\begin{split}
    &H_{\rm Z_2DSL}=\sum_{\langle{}ij\rangle}\chi_1\left(f^\dagger_{i,s}f^{}_{j,s}+f^\dagger_{j,s}f^{}_{i,s}\right) \\ 
    &+\sum_{i}\left[\eta_{1}\left(f^\dagger_{i,\uparrow}f^\dagger_{i+\hat{x},\downarrow}+f^\dagger_{i+\hat{x},\uparrow}f^\dagger_{i,\downarrow}\right)+{\rm H.c.}\right]\\
    &-\sum_{i}\eta_{1}\left[\left(f^\dagger_{i,\uparrow}f^\dagger_{i+\hat{y},\downarrow}+f^\dagger_{i+\hat{y},\uparrow}f^\dagger_{i,\downarrow}\right)+{\rm H.c.}\right]\\
    &+\sum_{i}\left[\eta_{5}\left(f^\dagger_{i,\uparrow}f^\dagger_{i+2\hat{x}+2\hat{y},\downarrow}+f^\dagger_{i+2\hat{x}+2\hat{y},\uparrow}f^\dagger_{+i,\downarrow}\right)+{\rm H.c.}\right]\\
    &-\sum_{i}\left[\eta_{5}\left(f^\dagger_{i,\uparrow}f^\dagger_{i-2\hat{x}+2\hat{y},\downarrow}+f^\dagger_{i-2\hat{x}+2\hat{y},\uparrow}f^\dagger_{+i,\downarrow}\right)+{\rm H.c.}\right],\label{eq:epHZ2}
\end{split}
\end{equation}
where $\chi_1$, $\eta_{1}$, and $\eta_{5}$ are variational parameters. Note that $\eta_{1}$ on the 1st NN bond and  $\eta_{5}$ on the 5th NN bond satisfy the $d_{x^2-y^2}$ and $d_{xy}$ symmetries of $D_{4}$ group, respectively. After the Fourier transformation, $H_{\rm Z_2DSL}$ becomes

\begin{equation}
\begin{split}
&H_{\rm Z_2DSL}=\left( f^\dagger_{{\bf k},\uparrow}~ f^\dagger_{{\bf k},\downarrow}~ f^{}_{-{\bf k},\uparrow}~ f^{}_{-{\bf k},\downarrow} \right) h_{\rm Z_2DSL}({\bf k})
\left(
\begin{array}{c}
f^{}_{{\bf k},\uparrow} \\ f^{}_{{\bf k},\downarrow} \\ f^\dagger_{-{\bf k},\uparrow} \\ f^\dagger_{-{\bf k},\downarrow} 
\end{array}
\right),\\
&h_{\rm Z_2DSL}=\left(
\begin{array}{cccc}
S_1 &  &  & A_1 + A_5 \\
 & S_1 & - A_1 - A_5 &  \\
 & - A_1 - A_5 & -S_1 &  \\
A_1 + A_5 &  &  & -S_1 \\
\end{array}
\right).
\end{split}
\end{equation}
Here
$$S_1({\bf k})=\chi_1(\cos k_x + \cos k_y),$$
$$A_1({\bf k})=\eta_1(\cos k_x - \cos k_y),$$
$$A_5({\bf k})=\eta_5\left[\cos (2k_x+2k_y) - \cos (2k_x-2k_y)\right].$$
At ${\bf k}=(\pi/2,\pm\pi/2)$, we find that 
$S_1({\bf k})=A_1({\bf k})=A_5({\bf k})=0$. Consequently, $H_{\rm Z_2DSL}$ exhibits Dirac cones at $(\pi/2,\pi/2)$ and $(\pi/2,-\pi/2).$ A $k\cdot p$ expansion around the Dirac cone of $(\pi/2,\pm\pi/2)$ leads to 
\begin{small}
\begin{equation*}
\begin{split}
\tilde{h}_{\rm Z_2}=&
\left(\begin{array}{c}
f^{\dagger}_{{\bf p},\uparrow} \\ f_{-{\bf p},\downarrow} 
\end{array}\right)^T
\left(\begin{array}{cc}
  \chi_1(-p_x\pm p_y)    &  \eta_1(-p_x\mp p_y) \mp 8\eta_5p_xp_y  \\
 h.c.    &  - \chi_1(-p_x \pm p_y)
\end{array}\right)
\left(\begin{array}{c}
f^{}_{{\bf p},\uparrow} \\ f^\dagger_{-{\bf p},\downarrow} 
\end{array}
\right)\\
&+\left(\begin{array}{c}
f^{\dagger}_{{\bf p},\downarrow} \\ f_{-{\bf p},\uparrow} 
\end{array}\right)^T
\left(\begin{array}{cc}
  \chi_1(-p_x\pm p_y)    &  \eta_1(p_x\pm p_y) \pm 8\eta_5p_xp_y  \\
 h.c.    &  - \chi_1(-p_x \pm p_y)
\end{array}\right)
\left(\begin{array}{c}
f^{}_{{\bf p},\downarrow} \\ f^\dagger_{-{\bf p},\uparrow} 
\end{array}
\right).
\end{split}
\end{equation*}
\end{small}

In Table~\ref{tabS:F}, we list the wave function fidelity between Gutzwiller projected Z$_2$DSL state and DMRG state on various cylinders.

\begin{table}
\caption{The wave function fidelity between Gutzwiller projected Z$_2$DSL state and DMRG state, $F=|\langle\Psi_{\rm Z_2DSL}|\Psi_{\rm DMRG}\rangle|$, on various cylinders. }\label{tabS:F}
    \begin{tabular}{  p{2.8cm} p{1.5cm} p{1.5cm} }
		\hline
		\hline
		\rule{0pt}{2.5ex}    
         & $F$ & $F^{1/L_x/L_y}$ \\ 
		\hline
       $L_y=4$,~~$L_x=12$  &  0.9367  &  0.9986 \\  
       $L_y=6$,~~$L_x=12$  &  0.9089  &  0.9987 \\
       $L_y=8$,~~ $L_x=8$   &  0.9149  &  0.9986 \\
       $L_y=8$,~~ $L_x=12$  &  0.8691  &  0.9985 \\
		\hline
		\hline
    \end{tabular}
\end{table}

Another possible QSL candidate is a $\pi$-flux state with the Hamiltonian as 
\begin{equation}
\begin{split}
H_{SU2-\pi}=&\sum_{i,s}\left[(-1)^{i_x}\left(f^\dagger_{i,s}f^{}_{i+\hat{y},s}+f^\dagger_{i+\hat{y},s}f^{}_{i,s}\right)\right]+\\
&\sum_{i,s}\left[\left(f^\dagger_{i,s}f^{}_{i+\hat{x},s}+f^\dagger_{i+\hat{x},s}f^{}_{i,s}\right)\right].    
\end{split}
\end{equation}
By noting that the only nonzero gauge flux on the elementary square plaquette expresses as $F\sim{}-\sigma^0$, the mean-field ansatz $H_{SU2-\pi}$ describes a SU(2) QSL with Dirac-type spinon excitations. We use SU(2)-$\pi$DSL to denote this state. However, the overlap between SU(2)-$\pi$DSL parton state and the DMRG state is almost zero, which indicates that even though SU(2)-$\pi$DSL also manifests Dirac cones, such an ansatz cannot efficiently describe the low-energy physics of the $J_1$-$J_2$ model.

\section{Details on the $J_1$-$J_2$-$J_{\chi}$ model}

The parity symmetry of the square lattice is broken when $J_\chi$ is finite. Then some PSG constraints in Eqs.~\eqref{eqS:PSGPyZ2DSL} and \eqref{eqS:PSGPxyZ2DSL} are removed. Consequently, the time-reversal-symmetry breaking terms with a general form of $i\chi_{n}^{\prime}\sigma^0+\eta_{n}\sigma^y$ are allowed. However, the remaining $C_4$ rotation symmetry requires that 
\begin{equation}
\begin{split}
&\mu_{i,i+\hat{x}}=\mathcal{G}[C_4, \mu_{i,i+\hat{x}}]\overset{!}{=}\sigma^z(\mu_{i,i+\hat{y}})^\dagger\sigma^z,\\
&\mu_{i,i+\hat{x}+\hat{y}}=\mathcal{G}[C_4, i+\hat{x}+\hat{y}]\overset{!}{=}\sigma^z\mu_{i,i+\hat{x}-\hat{y}}\sigma^z.
\end{split}
\end{equation}
The above constraints give rise to the CSL ansatz shown in Eq.~(5) in the main text.

\begin{figure}
	\includegraphics[width=0.75\linewidth]{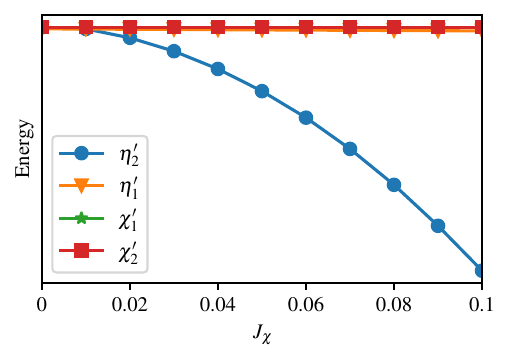}
	\caption{The variational ground-state energies are obtained by varying parton parameters shown in the legend. The other parton parameters are fixed as $\chi_1=1$, $\eta_1=1.42$, and $\eta_5=0.42$. The calculations are performed on a YC6 cylinder with $L_x=8$.}\label{figS:CSLpara}
\end{figure}

The CSL ansatz leads to four parton parameters, namely, $\chi_1'$, $\chi_2'$, $\eta_1'$, and, $\eta_5'$.
To determine the critical parameters of the $J_1$-$J_2$-$J_\chi$ model, we perform calculations of the variational energy by activating one of the four parameters at a time while setting the other three to zero. The results are shown in Fig.~\ref{figS:CSLpara}, which clearly demonstrates that only $\eta_2'$, the $id_{xy}$ pairing term on the 2nd NN bonds, significantly contributes to the variational energy.

Focusing on $\eta_2'$, the parton mean-field Hamiltonian becomes 
\begin{equation}
h_{\rm CSL}=h_{\rm Z_2DSL}+\left(
\begin{array}{cccc}
0 &  &  & iA_2 \\
 & 0 & -iA_2 &  \\
 & iA_2 & 0 &  \\
-iA_2 &  &  & 0 \\
\end{array}
\right),
\end{equation}
where 
$$A_2=\eta_2'\left[\cos(k_x+k_y)-\cos(k_x-k_y)\right].$$
To see the possible boundary zero modes, we perform the Fourier transformation only along the $\hat{y}$-direction. After simple calculations, we obtain a mean-field Hamiltonian in the mixed basis $f_{x,k_y,s}$ as 
\begin{equation*}
    \begin{split}
        H_{\rm CSL}=&\frac{\chi_1}{2}\sum_{x,k_y,s}\left( \cos k_yf^\dagger_{x,k_y,s}f_{x,k_y,s}+ f^\dagger_{x,k_y,s}f_{x+1,k_y,s}\right)+\\
        -&\frac{\chi_1}{2}\sum_{x,k_y,s}\left( \cos k_y f_{x,-k_y,s}f^\dagger_{x,-k_y,s}+ f_{x+1,-k_y,s}f^\dagger_{x,-k_y,s}\right)+\\
        &\frac{\eta_1}{2}\sum_{x,k_y}\left(f^\dagger_{x,k_y,\uparrow}f^\dagger_{x+1,-k_y,\downarrow}+f^\dagger_{x+1,k_y,\uparrow}f^\dagger_{x,-k_y,\downarrow}\right)+\\
        -&\frac{\eta_1}{2}\sum_{x,k_y}\left(f^\dagger_{x,k_y,\downarrow}f^\dagger_{x+1,-k_y,\uparrow}+f^\dagger_{x+1,k_y,\downarrow}f^\dagger_{x,-k_y,\uparrow}\right)+\\
        -&\eta_1\sum_{x,k_y}\cos{k_y}\left(f^\dagger_{x,k_y,\uparrow}f^\dagger_{x,-k_y,\downarrow}-f^\dagger_{x,k_y,\downarrow}f^\dagger_{x,-k_y,\uparrow}\right)+\\        i&\eta_5\sum_{x,k_y}\left[\sin2k_y\left(f^\dagger_{x,k_y,\uparrow}f^\dagger_{x+2,-k_y,\downarrow}-f^\dagger_{x+2,k_y,\uparrow}f^\dagger_{x,-k_y,\downarrow}\right)\right]+\\        -i&\eta_5\sum_{x,k_y}\left[\sin2k_y\left(f^\dagger_{x,k_y,\downarrow}f^\dagger_{x+2,-k_y,\uparrow}-f^\dagger_{x+2,k_y,\downarrow}f^\dagger_{x,-k_y,\uparrow}\right)\right]+\\        &\eta_2'\sum_{x,k_y}\left[\sin{}k_y\left(f^\dagger_{x,k_y,\uparrow}f^\dagger_{x+1,-k_y,\downarrow}-f^\dagger_{x+1,k_y,\uparrow}f^\dagger_{x,-k_y,\downarrow}\right)\right]+\\        -&\eta_2'\sum_{x,k_y}\left[\sin{}k_y\left(f^\dagger_{x,k_y,\downarrow}f^\dagger_{x+1,-k_y,\uparrow}-f^\dagger_{x+1,k_y,\downarrow}f^\dagger_{x,-k_y,\uparrow}\right)\right]+{\rm h.c.}.
    \end{split}
\end{equation*}
The above Hamiltonian exhibits zero modes at $k_y=\pm\pi/2$. Defining $a^\dagger_{x,s}=f^\dagger_{x,\pi/2,s}$ and $b^\dagger_{x,s}=f^\dagger_{x,-\pi/2,s}$, the $\pi/2$ sector of $H_{\rm CSL}$ reads
\begin{equation*}
\begin{split}
h^{\pi/2}_{\rm CSL}=&\frac{\chi_1}{2}\sum_{x,s}\left( a^\dagger_{x,k_y,s}a_{x+1,s}-b_{x+1,s}b^\dagger_{x,s}\right)+\\
    &\sum_{x,s}\theta_s\left[\left(\frac{\eta_1}{2}+\eta_2'\right)a^\dagger_{x,s}b^\dagger_{x+1,-s}+\left(\frac{\eta_1}{2}-\eta_2'\right)a^\dagger_{x+1,s}b^\dagger_{x,-s}\right]+\\
    &{\rm h.c.},
    \end{split}
\end{equation*}
where $-s$ is the inverse of spin $s$ index and  $\theta_{\uparrow(\downarrow)}=1(-1)$. 
Note that $h^{\pi/2}_{\rm CSL}$ is just the Kitaev chain when $\eta_1=0$. Moreover, it exhibits exact zero modes when $\eta_2'=\chi_1/2.$ However, the system also exhibits nodal points at ${\bf k}=(0,\pi)$ and ${\bf k}=(\pi,0)$, which is not a gapped chiral spin liquid. In this case, this parton Hamiltonian supports boundary zero modes for every $k_y$.
When $\eta_1>0$, the system supports two Majorana zero modes on the boundaries of the effective 1D chain for each spin index $s$ and each momentum $k_y=\pm\pi/2$.

\section{More details on DMRG data}

To analyze the convergence of our DMRG simulations, we study the scaling behavior of the ground-state energy with the DMRG truncation errors.
As shown in Fig.~\ref{figS:scaling},  the ground-state energy scale linearly with DMRG truncation errors for bond dimensions $\chi_{D}\geq{}4000$, suggesting a faithful convergence in our DMRG simulations. 

\begin{figure}
	\includegraphics[width=1.\linewidth]{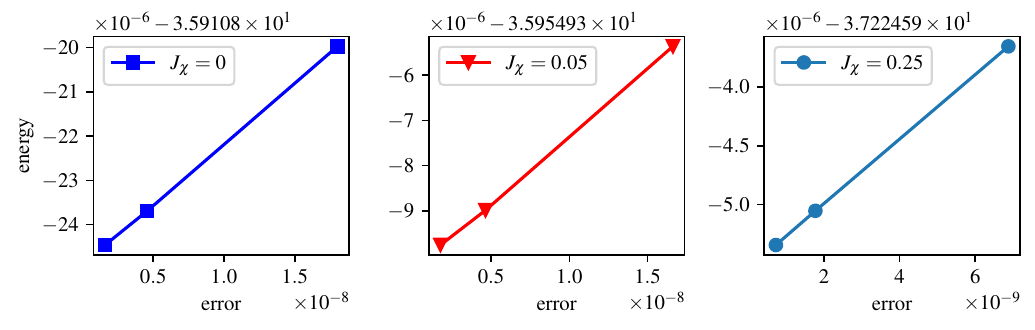}
	\caption{Ground-state energies of the $J_1$-$J_2$-$J_\chi$ model as functions of DMRG truncation errors. The calculations are performed on YC6 cylinders with $L_x=12$.
	}\label{figS:scaling}
\end{figure}

\begin{figure}
	\includegraphics[width=0.8\linewidth]{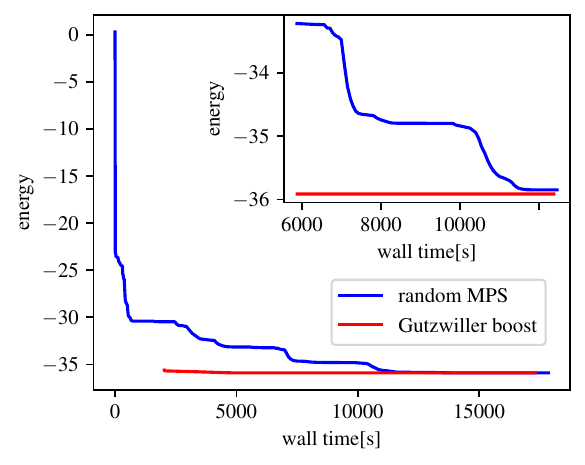}
	\caption{The ground-state energy for a specific DMRG sweep with the total wall time cost after the sweep. 
    The time cost of preparing a Z$_2$DSL parton state at bond dimension$\chi_S=2000$ is about 2000s. The inset shows a zoom-in plot between an interval of [$\sim{}6000$s, $\sim{}12000$s]. 
    The calculations are performed on a YC6 cylinder with $L_x=12$ and $\chi_D=4000$.}\label{figS:walltime}
\end{figure}
To see the quality of our Z$_2$DSL parton ansatz, we perform DMRG simulations initialized with (i) a randomly generated product state and (ii) a Gutzwiller projected parton state, respectively. Fig.~\ref{figS:walltime} shows how DMRG calculations converge with different initial states. Compared to traditional DMRG, the Gutzwiller-Boosted DMRG can shorten the convergence time to one-third. For example, the energy calculation using Gutzwiller-Boosted DMRG reaches a plateau at a wall time of approximately 6000 seconds. In contrast, conventional DMRG requires about 18000 seconds to achieve the same energy.

\section{Instability against VBS state and N\'eel order}

{\color{black}
The $J_1$-$J_2$ model, as a highly frustrated system, hosts intense energetic competition between possible ground states. For instance, in addition to the QSL phase, there might exist a valence bond solid (VBS) phase between the N\'eel and stripe orders. This raises a natural question: How is the Z$_2$ Dirac QSL stable against the instability of a VBS state?

\begin{figure}
    \centering
\includegraphics[width=1.0\linewidth]{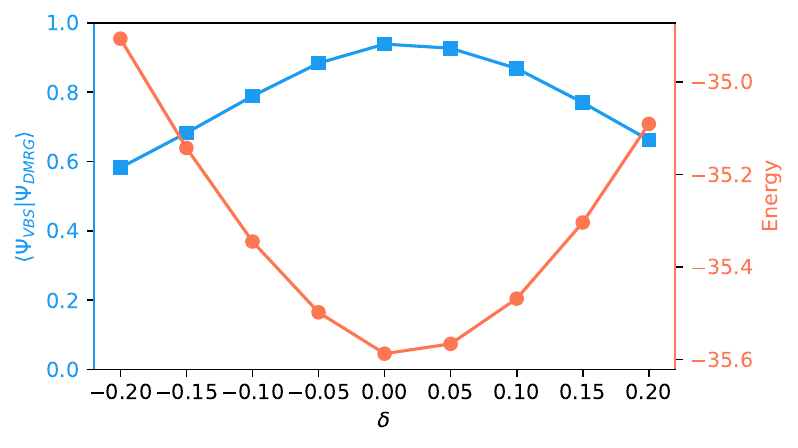}
    \caption{Left axis: the wave function fidelity between $|\Psi_{\rm VBS}\rangle$ and $|\Psi_{\rm DMRG}\rangle$ as function of stripy index $\delta$. Right axis: the variational energy of the VBS parton state at $J_2=0.5$. The calculations are carried out on a $L_y\times{}L_x=6\times12$ cylinder with a VBS state of $(\eta_1,\eta_5)=(1.65,0.55)$.
}
    \label{fig:VBS_delta}
\end{figure}

We begin with constructing such a the VBS state. Introducing a stripy index $\delta$ which breaks the translational symmetry along the $x$ direction, the NN hopping $\chi$ are modulated as
\begin{equation}
\begin{split}
    &\chi_{i,i+\hat{x}}=\left[1-(-1)^{i_x}\delta\right]\left[\chi\sigma^z+\eta_1\sigma^x\right],\\
    &\chi_{i,i+\hat{y}}=\chi\sigma^z+\eta_1\sigma^x 
\end{split}
\end{equation}
where $i_x$ is the $x$-coordinate of site $i$. As shown in Fig.~\ref{fig:VBS_delta}, we measure the energy of this VBS parton state $|\Psi_{\text{VBS}}\rangle$ at the point of $J_2=0.5$, as well as the fidelity between $|\Psi_{\text{VBS}}\rangle$ and $|\Psi_{\text{DMRG}}\rangle$. 
The optimal energy and fidelity are consistently achieved at $\delta^{*}=0$, indicating the absence of a VBS phase at $J_2=0.5$. 

Another piece of evidence is the value of the optimal $\eta_2$, the pairing parameter on the 2$^{\text{nd}}$ NN bonds. The VBS phase emerges due to the tight energetic competition between $J_1$ and $J_2$. Therefore, if there is a VBS state at $J_2 = 0.5$, the amplitudes of mean-field parameters on the 2$^{\text{nd}}$ NN bonds should be comparable to those on the 1$^{\text{st}}$ NN bonds. However, our variational results clearly show that the optimal value of $\eta_2$ is as small as $0.04$ at $J_2 = 0.5$, while the optimal value of $\eta_1$ is as large as $1.5$. This is in stark contrast with the argument that the VBS phase is an instability of the Z$_2$ Dirac QSL phase.

\begin{figure}
    \centering
    \includegraphics[width=0.85\linewidth]{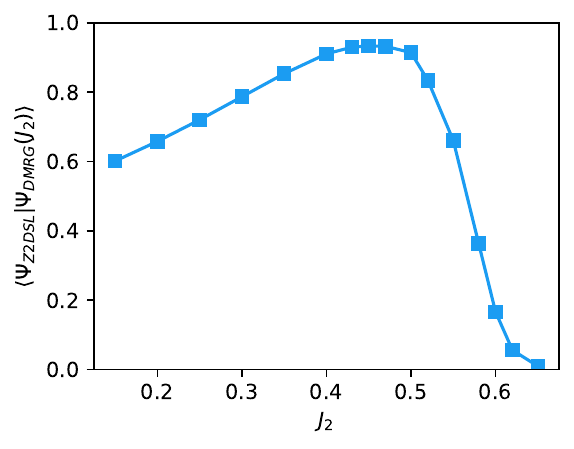}
    \caption{The wavefunction fidelity as a function of $J_2$. Here the Z$_2$DSL parton state is prepared with parameters $(\eta_1,\eta_5)=(1.65,0.55)$ at bond dimension $D=2000$, and the ground state is obtained by Gutzwiller-guided DMRG.}
    \label{fig:fidelity_J2}
\end{figure}

We further evaluating the fidelity as a function of $J_2$ for the Z$_2$DSL parton ansatz. As shown in Fig.~\ref{fig:fidelity_J2}, we observe that in the QSL phase, the fidelity remains above 0.9 on a $6\times{}12$ cylinder, indicating that the whole QSL phase can be efficiently described by our ansatz. Note that the maximum fidelity $F\approx{}0.93$ is reached at $J_2=0.47$. When $J_2>0.6$, the fidelity is almost zero, indicating an emergence of VBS or stripe phase. 
In the N\'eel ordered phase, the fidelity is still considerable. However, it becomes larger when we turn off $\eta_1$ and $\eta_5$, namely, the parameters introducing Z$_2$ gauge fluctuations. This suggests that the N\'eel phase perhaps is a U(1) RVB in which the long-ranged magnetic order is achieved due to the confinement of the U(1) gauge field. 

\begin{figure}
    \centering
\includegraphics[width=0.85\linewidth]{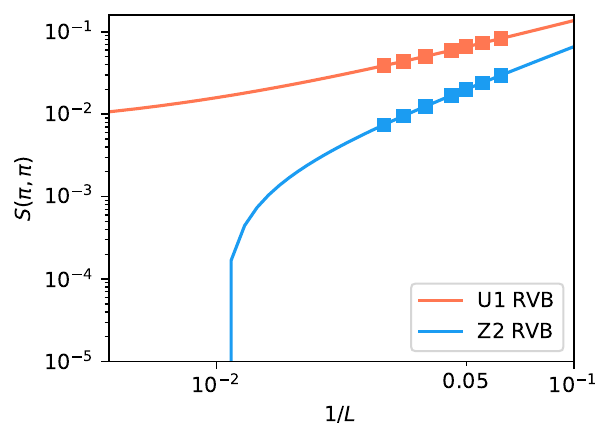}
    \caption{The spin structure factor $S(q=\pi,\pi)$ of  the U(1) and Z2 RVB states on a $L\times L$ square lattice as a function of $1/L$,  the inverse linear scale of the lattice up to $L=34$. The U(1) RVB is given by  $\eta_1=0.15$ and  $\eta_2=\eta_5=0$, as well as the Z$_2$ RVB is given by  $\eta_1=1.5$, $\eta_2=0.1$, and $\eta_5=0.5$. The errorbar is much less than the size of markers. The solids lines are fitted by a function of $bL^{-a}+c$, where the fitted $c\sim{}0.01$ for U(1) RVB and $c\sim{}-10^{-8}$ for Z$_2$ RVB. }
    \label{fig:Szqq}
\end{figure}

In order to verify this, we optimize the Z$_2$DSL ansatz for the Hamiltonian at $J_2=0$.
We find that the effective wave function can be described by a U(1) RVB state which exactly preserves the SU(2) spin rotational symmetry. 
The per-site wave function fidelity between this U(1) RVB state and DMRG-optimized state is as high as $0.994$~\cite{Jin2025}. In the framework of our Z$_2$DSL ansatz, this U(1) RVB can be roughly parameterized as
\begin{equation}
\chi=1,  \quad  \eta_1=0.15, \quad    \eta_2=\eta_5=0,    
\end{equation} 
where the U(1) gauge fluctuations are recovered due to the absence of $\eta_2$ and $\eta_5$. Using large-scale variational Monte Carlo method, we find that the static spin structure factor $S(q=\pi,\pi)$ of this U(1) RVB state remains a finite value in the thermodynamic limit, as shown in Fig.~\ref{fig:Szqq}. This important result indicates that the phase transition between N\'eel order and Dirac QSL occurs when the $Z_2$ fluctuations emerge with increasing $J_2$, which may allow us to write down an effective field theory for this phase transition in future studies.  

}

\section{Topological ground-state degeneracy}

The ground-state degeneracy manifests the non-trivial ground-state structure. For a topological order, its ground-state degeneracy depends on the specific geometry of the underlying lattice. Here, to investigate the difference between Gutzwiller projected Z$_2$ and U(1) CSLs, we study the corresponding topological degeneracy on toroidal geometries.

For a general parton theory on torus, there are two distinct global fluxes, $\Phi_x$ and $\Phi_y$, along the $\hat{x}$ and $\hat{y}$ directions, respectively. Overall, there are four distinct choices of those global fluxes, denoted by $(\Phi_x,\Phi_y)$. Here $\Phi_c=0$ ($c=x,y$) for the periodic boundary condition along the $\hat{c}$ direction and $\Phi_c=\pi$ for the antiperiodic boundary condition. Therefore, applying the Gutzwiller projection, we obtain four spin states denoted by $|\Phi_x,\Phi_y\rangle$ (e.g., $|0,0\rangle$, $|\pi,0\rangle$, and $|\pi,\pi\rangle$).

We then calculate the overlaps between all possible states on a $L_x=L_y=L$ torus, leading to a $4\times{}4$ overlap matrix. For small system sizes ($L=4,6$), we convert the Gutzwiller project state into MPSs and then calculate the overlaps directly (up to a truncation error due to the finite bond dimensions of MPS). For a larger lattice size up to $L=16$, we use the variational Monte Carlo method to compute the overlap. Note that when using variational Monte Carlo, we first calculate the square of overlap 
\begin{eqnarray}
    \frac{\langle\Psi_a|\Psi_b\rangle\langle\Psi_a|\Psi_b\rangle}{\langle\Psi_a|\Psi_a\rangle\langle\Psi_b|\Psi_b\rangle}.\label{eq:SqrOvlp}
\end{eqnarray}
Then, the value of overlap can be obtained by taking a square root of Eq.~\eqref{eq:SqrOvlp}. The additional Z$_2$ number, which is caused by the square root, is determined by the sign of $\langle\Psi_a|\Psi_b\rangle/\langle\Psi_a|\Psi_a\rangle$.

The linearly independent ground states can be constructed by diagonalizing the $4\times{}4$ overlap matrix. For the U(1) CSL with $\eta_2=\eta_5=0$ and $\eta_2'>0$, we find that there are only two non-zero eigenvalues of the overlap matrix, indicating a two-fold degeneracy for the U(1) CSL. For the Z$_2$ CSL with finite $\eta_2$ and $\eta_5$, we find that all of those four eigenvalues are positive. Nevertheless, two of those four eigenvalues are relatively small and highly depend on the values of $\eta_2$ and $\eta_5$. To suppress the finite-size effects, we tune (in the unit of $\chi_1=1$) $\eta_2=0.1$ and $\eta_5=0.6$ to enhance the Z$_2$ fluctuations. We also set $\eta_2'=0.6$ to ensure a sufficiently large chiral gap. We observe that these two smaller eigenvalues are substantially finite and exhibit no trend of decrease as the system size $L$ grows. For instance, the two small eigenvalues are about $0.085\pm0.002$ and $0.095\pm0.002$, respectively, for $L=16$. Therefore, the Z$_2$ CSL has four-fold topologically degenerate ground states on tori.

\section{Field theory of criticality between U(1) and Z$_2$ CSLs}
{\color{black}
Here we provide a field theory for the quantum phase transition between the U(1) and Z$_2$ CSL. We start from the U(1) CSL at a relatively large $J_\chi$. The parton mean field ansatz is a Chern insulator with Chern number $c=1$ for each spin. We argue that the phase transition is driven by the onset of $\eta_5$, which higgses the U(1) gauge field to a Z$_2$ one. 

In field theory, the fluctuations of $\eta_5$ can be represented by a complex scalar field (boson) $\Phi$ which enters the mean field ansatz in the form of $\Phi (\epsilon_{\sigma \sigma'} f^\dagger_{\sigma}f^\dagger_{\sigma'}+h.c.)$. Note that in both U(1) and Z$_2$ CSL phases, the spinon fields $f_\sigma$ are gapped across the transition and do not enter the low energy field theory.
The Lagrangian is then in the form of an Abelian Higgs theory:
\begin{align}
\mathcal L&=|(\partial_\mu - 2ia_\mu)|\Phi|^2- s|\Phi|^2-g|\Phi|^4 \notag \\
&~~~-\frac{1}{4\pi}\alpha_1 {\rm d} \alpha_1-\frac{1}{4\pi}\alpha_2 {\rm d} \alpha_2 \notag \\ 
&~~~+ \frac{1}{2\pi}(\frac{1}{2}A_s-a){\rm d}\alpha_1+\frac{1}{2\pi}(-\frac{1}{2}A_s-a){\rm d}\alpha_2
\end{align}
where $a{\rm d}b$ is an abbreviation of the Chern-Simons (CS) term $\epsilon_{\mu \nu \sigma}a_\mu \partial_\nu b_\sigma$, and $a_\mu$ ($\mu=1,2$) denote two U(1) gauge fields which couples to the fermionic spinon field $f_\sigma$. Note that our condensation of $\Phi$ is a Cooper pair of spinons and thus carries charge $2$ under the gauge fields $a_\mu$. Because we consider a spinful theory, the fields $\alpha_1$ and $\alpha_2$ are two dual U(1) gauge fields to capture the $c=1$ Chern insulator populated from $f_\uparrow$ and $f_\downarrow$. $A_s$ is a probing U(1) field corresponding to total $S_z$ conservation. We introduce $A_s$ to capture the spin quantum Hall effect of CSL.

The phase transition is effectively tuned by $s$, the mass term of the boson field $\Phi$. When $s>0$, $\Phi$ is gapped and we are left with the Lagrangian:
\begin{align}
    \mathcal L_{\mathrm{U(1) -CSL}}&=-\frac{1}{4\pi}\alpha_1 {\rm d} \alpha_1-\frac{1}{4\pi}\alpha_2 {\rm d} \alpha_2 \notag \\ 
&~~~+ \frac{1}{2\pi}(\frac{1}{2}A_s-a){\rm d}\alpha_1+\frac{1}{2\pi}(-\frac{1}{2}A_s-a){\rm d}\alpha_2
\end{align}
By integrating $a$, the dual U(1) fields are locked as $\alpha_1=-\alpha_2=\alpha$, leading to the final form of Lagrangian:
\begin{equation}
    \mathcal L_{\mathrm{U(1)-CSL}}=-\frac{2}{4\pi}\alpha {\rm d} \alpha+\frac{1}{2\pi} A_s {\rm d} \alpha.
\end{equation}
This is just the well-known low energy effective theory for the bosonic Laughlin state, which indicates that the U(1) CSL can be viewed as the $\nu=1/2$ bosonic Laughlin state.

On the other hand, when $s<0$, we have a condensation of $\Phi$ field, namely, $\langle \Phi \rangle \neq 0$, which higgses the U(1) gauge field $a_\mu$ down to a Z$_2$ gauge field. To derive an effective theory, we introduce a dual U(1) gauge field $b$ for the condensation $\langle \Phi \rangle$ and reach the effective theory for the Z$_2$ CSL:
\begin{align}
    \mathcal L_{\mathrm{Z_2 -CSL}}&=\frac{2}{2\pi}b {\rm d} a-\frac{1}{4\pi}\alpha_1 {\rm d} \alpha_1-\frac{1}{4\pi}\alpha_2 {\rm d} \alpha_2 \notag \\ 
&~~~+ \frac{1}{2\pi}(\frac{1}{2}A_s-a){\rm d}\alpha_1+\frac{1}{2\pi}(-\frac{1}{2}A_s-a){\rm d}\alpha_2
\end{align}
Now integrating the field $a$ locks all the U(1) gauge fields as $\alpha_1+\alpha_2=2b$. We can then express the Lagrangian in terms of $\alpha_1$ and $b$ by substituting $\alpha_2=2b-\alpha_1$.  For simplicity, we make another linear combination: $\alpha_1=-\alpha+\beta$ and $b=\beta$. In the end, we substitute $\alpha_1=-\alpha+\beta$ and $\alpha_2=\alpha+\beta$ and reach the final theory:
\begin{align}
    \mathcal L_{\mathrm{Z_2 -CSL}}&=-\frac{2}{4\pi}\alpha {\rm d} \alpha-\frac{1}{2\pi}A_s {\rm d} \alpha -\frac{2}{4\pi}\beta {\rm d} \beta.
\end{align}

One can see that it contains two decoupled $U(1)_2$ theories. So the total number of anyons is $2\times 2=4$.
At the critical point with $s=0$, we can also simplify the critical theory by integrating $\alpha_1$ and  $\alpha_2$. The critical theory is:
\begin{align}
    \mathcal L_{\mathrm{critical}}=|(\partial_\mu-i 2 a_\mu)|\Phi|^2- s|\Phi|^2-g|\Phi|^4+\frac{2}{4\pi} ada+\frac{1}{8\pi}A_s {\rm d} A_s
\end{align}
The critical theory is in the form of Abelian Higgs theory with a Chern-Simons term. The last term, $\frac{1}{8\pi}A_s {\rm d} A_s$, is just a background fractional spin quantum Hall effect which does not change across the transition. Also across the transition the spinon remains gapped.  The critical fluctuation happens only in the spin singlet sector. This means that even at the quantum critical point, we expect spin-spin correlation function to be exponentially decayed just like a gapped spin liquid. In neutron scattering experiments, the spectrum also remains gapped across the transition. On the other hand, there may be a power-law correlation function for the spin chirality operator $\vec S_i \cdot (\vec S_j \times \vec S_k) \sim da$. But it may be quite challenging to detect this gapless mode in finite size numerical simulation.
}

\end{document}